\documentclass[12pt,times]{article} 
\usepackage{graphics,graphicx,epsfig,sidecap,amsmath,amssymb}
\sidecaptionvpos{figure}{t}
\usepackage{natbib} 
\usepackage{fancyhdr}
\include{journal}
\usepackage{setspace}
\usepackage{comment}
\usepackage[most]{tcolorbox}
\usepackage[colorlinks=true,linkcolor=blue,citecolor=blue]{hyperref}
\hypersetup{urlcolor=blue}

\definecolor{firebrick}{rgb}{0.7, 0.13, 0.13}

\setcitestyle{aysep={}}

\def\eps@scaling{.95}
\def\epsscale#1{\gdef\eps@scaling{#1}}

\def\plotone#1{\centering \leavevmodehttps://www.overleaf.com/project/6489dfa16d9030f468eb788b
    \epsfxsize=\eps@scaling\columnwidth \epsfbox{#1}}

\usepackage{color} 

\newcommand{\blue}[1]{\textcolor{blue}{#1}}
\newcommand{\fire}[1]{\textcolor{firebrick}{\textbf{#1}}}

\usepackage{color}

\newcommand{\apjl}{ApJL}
\newcommand{\apjs}{ApJS}

\newcommand{\aap}{A\&A}

\newcommand{\micron}{$\mu$m}

\newtcolorbox{gray}{boxrule=0pt, frame empty,
notitle,
colback=yellow!5!white
}

\usepackage{natbib}
\bibpunct{[}{]}{;}{n}{}{,}

\normalsize
\begin{document}

\begin{center}
    \Large
    \textbf{Roman CCS White Paper}
\end{center}

\begin{center}
    \Large
    \textbf{Discovery and Characterization of Galactic-scale Dual Supermassive Black Holes Across Cosmic Time}
\end{center}
\normalsize

\noindent {\bf Roman Core Community Survey:} High Latitude Wide Area Survey

\noindent {\bf Scientific Categories:} supermassive black holes and active galaxies

\noindent {\bf Additional Scientific Keywords:} High-luminosity active galactic nuclei, High-redshift galaxies, Interacting galaxies, Quasars, Supermassive black holes \\

\noindent \textbf{Submitting Author:}

Yue Shen, University of Illinois at Urbana-Champaign, shenyue@illinois.edu \\

\noindent \textbf{List of Contributing Authors:}


J. Andrew Casey-Clyde, University of Connecticut, andrew.casey-clyde@uconn.edu

Yu-Ching Chen, University of Illinois at Urbana-Champaign, ycchen@illinois.edu 

Arran Gross, University of Illinois at Urbana-Champaign, acgross@illinois.edu 

Melanie Habouzit, Zentrum f\"ur Astronomie \& Max-Planck-Institut (Germany), habouzit@mpia.de

Hsiang-Chih Hwang, Institute for Advanced Study, hchwang@ias.edu

Yuzo Ishikawa, Johns Hopkins University, yishika2@jhu.edu

Jun-Yao Li, University of Illinois at Urbana-Champaign, junyaoli@illinois.edu

Xin Liu, University of Illinois at Urbana-Champaign, xinliuxl@illinois.edu

Chiara M. F. Mingarelli, Yale University, chiara.mingarelli@yale.edu

D.\ Porquet, Aix Marseille Univ., CNRS, CNES, LAM, Marseille, France, delphine.porquet@lam.fr

Aaron Stemo, Vanderbilt University, aaron.m.stemo@vanderbilt.edu

Ming-Yang Zhuang, University of Illinois Urbana-Champaign, mingyang@illinois.edu

\newpage
\setcounter{page}{1}

\begin{gray}
\blue{The High Latitude Wide Area Survey with Roman, with its unprecedented combination of sensitivity, spatial resolution, area and NIR wavelength coverage, will revolutionize the study of galactic-scale environments of SMBH pairs. This white paper summarizes the science opportunities and technical requirements on the discovery and characterization of SMBH pairs down to galactic scales (i.e., less than tens of kpc) over broad ranges of redshift ($1<z<7$) and luminosity ($L_{\rm bol}\gtrsim 10^{42}\,{\rm erg\,s^{-1}}$). }
\end{gray}

The hierarchical structure formation paradigm predicts the formation of pairs of supermassive black holes in merging galaxies. When both (or one) members of the SMBH pair are unobscured AGNs, the system can be identified as a dual (or offset) AGN. Quantifying the abundance of these AGN pairs as functions of separation, redshift and host properties is crucial for understanding SMBH formation and AGN fueling in the broad context of galaxy formation. It is also a critical pathway to forecasting low-frequency (nHz-mHz) gravitational waves from the coalescence of SMBHs, whose detection is imminent with pulsar timing arrays and future space-based interferometers~\cite{Goulding_etal_2019, CaseyClyde2022, Mingarelli2022, Koss2023}. Following the merger of two galaxies, the two SMBHs within each of the galaxies will pair up and eventually evolve into a gravitationally bound binary, separated by $\lesssim10$\,parsec, via dynamical friction and other interactions with gas and stars \citep[e.g.,][]{Begelman_etal_1980,gould00,milosavljevic01,blaes02,Yu_2002,Khan2013,DEGN,Ogiya2022}. Studying SMBH pairs (via their proxy, AGN pairs) at different evolutionary stages, e.g., from $>$tens of kpc separations to $\lesssim$10\,pc scales of bound SMBH binaries, provides the foundation to understand the dynamical formation of binary SMBHs \citep[e.g.,][]{colpi09,Dotti2012,Volonteri2016}. 

The scales from tens of kpc to tens of parsec (i.e., galactic scales) are of particular interest. These scales correspond to the late stage of galaxy mergers, while the two SMBHs are on their way to become a bound binary\footnote{A separate Roman white paper is dedicated to bound binary massive black holes with the High Latitude Time Domain Survey \cite{Roman-WP-Haiman}.}. The observed frequency of pairs of AGNs on these galactic-scales is essential for testing physical recipes for AGN fueling in cosmological simulations and for constraining how fast the BH pair forms a bound binary \citep[e.g.,][]{yu11,Steinborn2015,Dosopoulou2017,Kelley2017,Tremmel2018}. In addition, the $\sim$\,kpc-scale SMBH pair frequency may provide observational constraints on the physical nature of dark matter particles. For example, in fuzzy dark matter \citep[][]{Hu2000}, a form of dark matter that consists of extremely light scalar particles with masses on the order of $\sim$10$^{-22}$ eV, SMBH pairs would never get much closer than $\sim$1 kpc because fuzzy dark matter fluctuations may inhibit the orbital decay and inspiral at kpc scales, resulting in a pile-up of $\sim$~kpc dual AGNs \citep{Hui2017}. \\

\noindent\fire{Current Status and Challenges}\quad Cosmological hydrodynamical simulations now have sufficient volume to probe dual/offset AGN populations at high redshift \citep{Blecha_etal_2016,DeRosa_etal_2019, Steinborn2016, ChenN_etal_2022,Volonteri_etal_2022}, where mergers occur much more frequently than in the low-redshift universe. These simulations have predicted populations of dual and offset AGNs from cosmic dawn to low redshift, along with detailed properties of their host galaxies. Despite these recent theoretical advances, there are still substantial variations in the modeling details in these simulations and the forecasts differ among different work. There is also a serious mismatch in the AGN luminosities from current simulations and observations, with the latter often limited to the most luminous quasar population ($L_{\rm bol}\gtrsim 10^{45}\,{\rm erg\,s^{-1}}$), which has poor statistics even in the largest simulation volume currently available \cite{ChenN_etal_2022,ChenN_etal_2023}.  

\begin{figure}[!h]
\centering
    \includegraphics[width=0.95\textwidth]{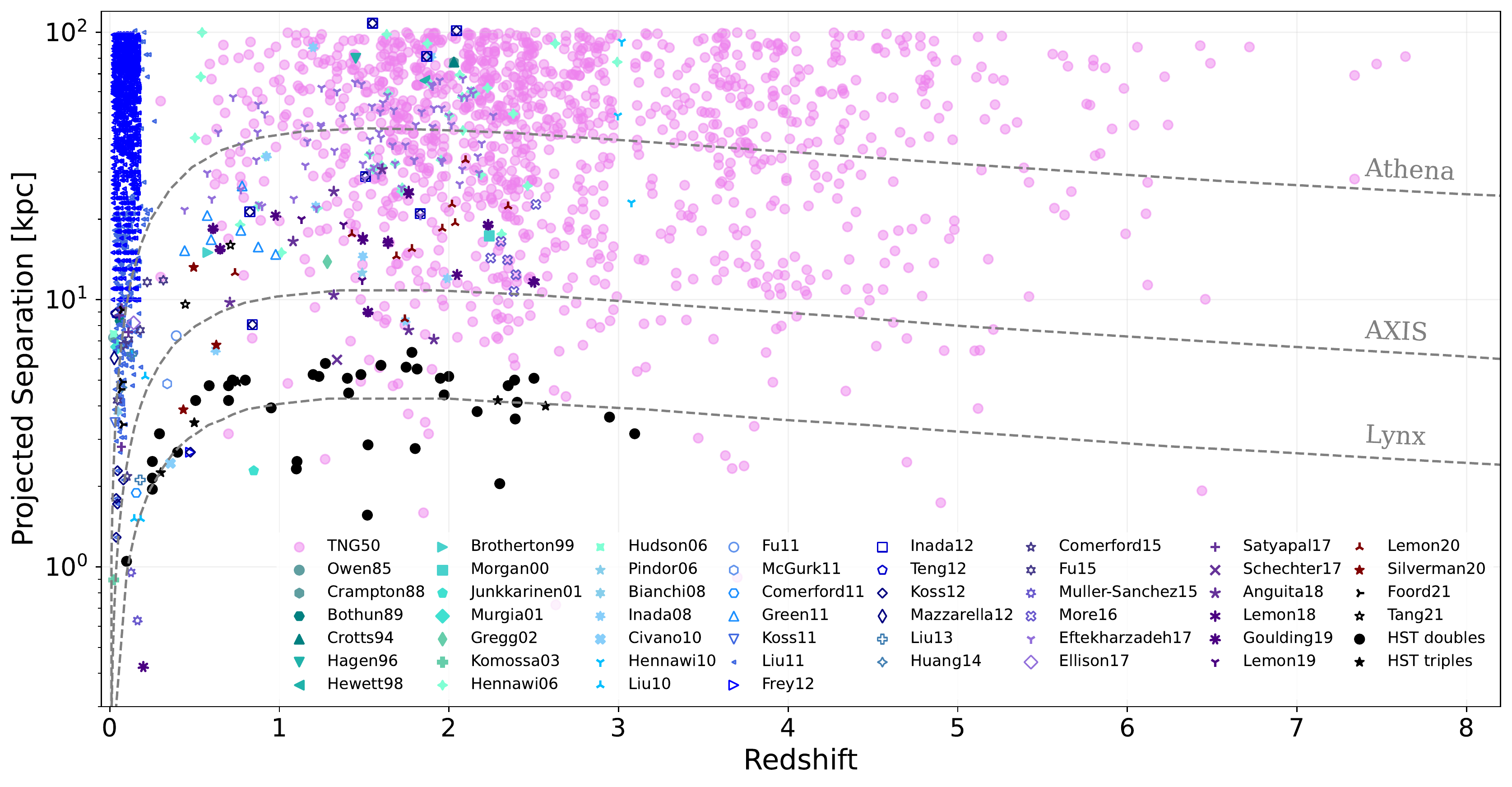}
    \caption{Current inventory of observed dual AGNs/quasars in the redshift-separation space, along with predicted dual AGNs from the TNG50 simulation (pink points). The black circles are {\em candidate} (a few are now confirmed) dual quasars selected using Gaia data. Confirmed dual AGNs at sub-arcsec separations are sparse at $z>1.5$. Figure adapted from \cite{Chen_etal_2022} and Puerto-Sanchez, Habouzit et al. in prep. \textbf{\em Roman will directly resolve pairs down to $\sim 0.1$" separations and discover these small-scale systems on industrial scales, and extend the search to redshifts of up to $z\sim 7$. } }
    \label{fig:z_sep}
\end{figure}

On the observational side, there is a stark paucity of robustly confirmed $\sim$kpc-scale dual or offset AGNs at $z>1$, as highlighted in Fig.~\ref{fig:z_sep}. The main observational hurdle is the stringent requirements for spatial resolution and robust AGN identification. Kpc-scale dual and offset AGN identification requires sub-arcsec resolution; the rareness of AGNs in pairs requires wide-area searches. So far, ground-based wide-area imaging and spectroscopic surveys have been highly inefficient in finding sub-arcsec dual/offset AGNs. 

The all-sky Gaia data have recently been used to select candidate sub-arcsec dual/offset AGNs and lenses at $z>1$ \cite[]{Hwang_etal_2020,Shen_etal_2021NatAs,Mannucci_etal_2022,Shen_etal_2023,Lemon_etal_2022}. Yet, follow-up observations, typically with HST resolution, are required to confirm the dual/offset AGN nature. It is a painstaking process to rule out alternative scenarios, i.e., a gravitationally lensed AGN instead of a genuine AGN pair \cite{Chen_etal_2022,Chen_etal_2023}. Steps from Gaia selection to HST/JWST follow-up for confirmation are time consuming. As a result, unambiguously confirmed dual quasars at $z>1.5$ are currently limited to a handful of objects \cite{Chen_etal_2023}. 

\begin{SCfigure}[][h!]
\centering
    \includegraphics[width=0.55\textwidth]{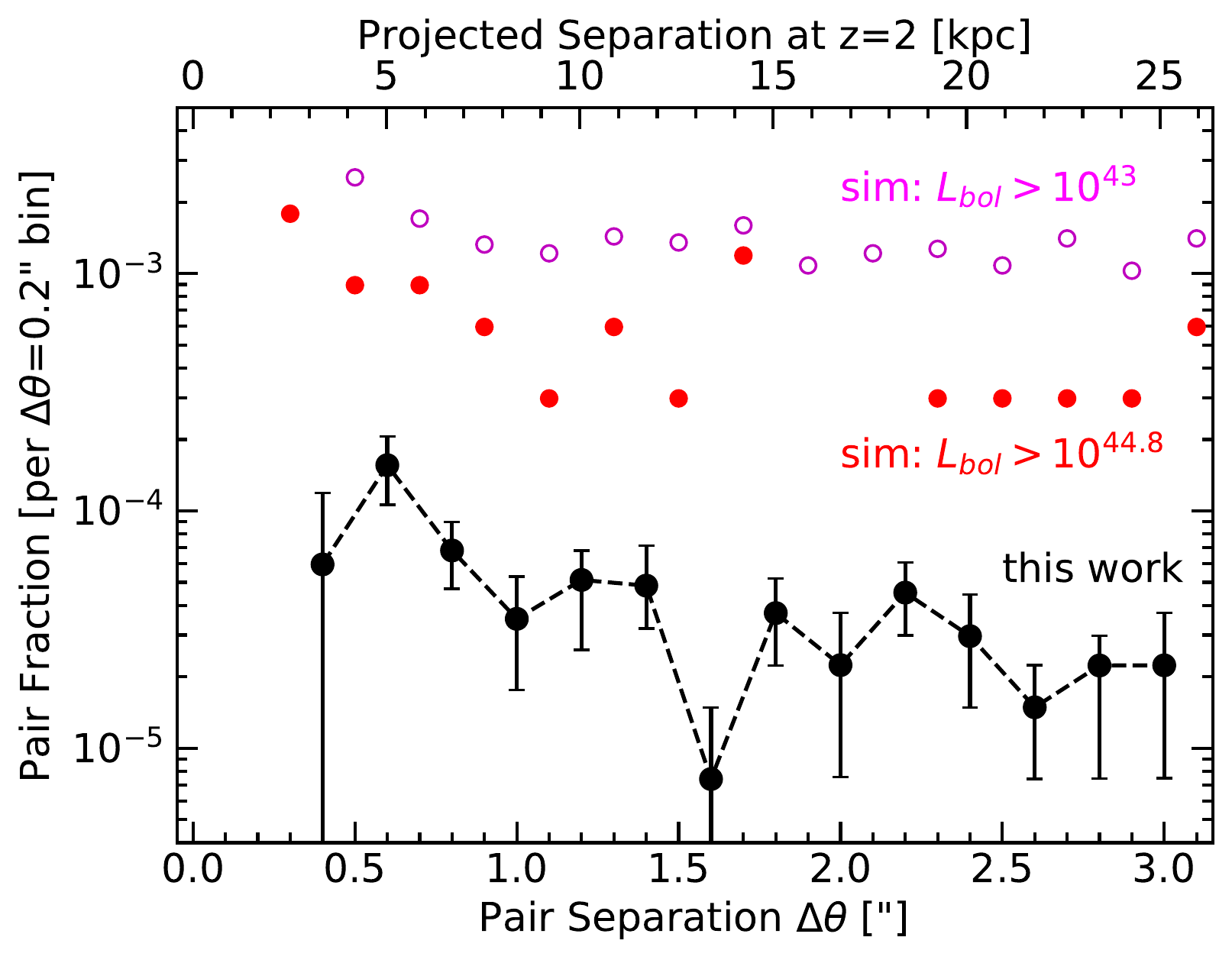}
    \caption{Constraints on the dual quasar fraction (among all quasars) as a function of pair separation (black points with error bars) using Gaia-resolved quasar pairs ($L_{\rm bol}> 10^{45.8}\,{\rm erg\,s^{-1}}$) \cite{Shen_etal_2023}. These constraints should be treated as upper limits as some of these double systems may be lensed quasars. The red and magenta circles are predicted dual AGN fractions for two luminosity thresholds, which are fainter than the observed quasar sample by at least one order of magnitude. Figure adapted from \cite{Shen_etal_2023}.}
    \label{fig:stat}
\end{SCfigure}

While existing statistics from Gaia are starting to constrain the galactic-scale dual quasar frequency as a function of separation (Fig.~\ref{fig:stat}), the mismatch between the AGN luminosities in observations and simulations prevents a fair comparison. It is anticipated that treasury extragalactic HST/JWST programs will push these searches to faint AGN luminosities \cite[e.g.,][]{Stemo_etal_2021}, but to ensure statistics, deep and wide area searches are necessary. \\ 

\noindent\fire{Enter Roman}\quad Without doubt, Roman will transform the landscape of dual/offset AGN research with its High Latitude Wide Area Survey. In order to significantly improve observed pair statistics, to extend to lower AGN luminosities matching simulations, and to explore the diversity in SMBH and host properties, it is necessary to carry out searches with deep, wide-area surveys at sub-arcsec resolution. Upcoming wide-field space missions, including Euclid \citep[][to be launched in $\sim 2023$]{Euclid_EWS}, the Chinese Space Station Telescope \citep[CSST,][to be launched in $\sim 2024$]{Zhan_2021}, and Roman \citep[][to be launched before $\sim 2027$]{Spergel_etal_2015}, will provide the perfect combination to conduct systematic searches of dual/offset AGNs across cosmic time. All three missions will carry out a wide-field imaging survey in multiple filters with $\sim 0.05-0.2$'' resolution and depths of $\sim 25-28$ AB mag, with additional spectroscopic capabilities. The combined photometric data cover a broad wavelength range across UV-optical-near-infrared. These data can be used to efficiently select candidate AGN pairs based on photometric colors and spectroscopic information, down to the diffraction limit of these space telescopes.

In particular, the deep and high-resolution IR imaging from Roman will provide the most crucial component to test the lensing scenario for high-redshift double quasars. Optical imaging with HST resolution is insensitive to detect a potential $z>1$ lens galaxy \cite[][]{Shen_etal_2021NatAs} and the solution is to use near-IR imaging \cite{Chen_etal_2023}; on the other hand, JWST would be too expensive to cover wide areas. In addition, the capability of detecting the host galaxy in deep IR imaging and measuring sub-arcsec offset of point sources within will enable the systematic discovery of single offset AGNs in high-redshift mergers. Host galaxy measurements will also allow a detailed look at the populations of dual and offset AGNs in different types of galaxies, shedding light on AGN fueling and recoiling SMBHs. With combined data sets from these upcoming space-based surveys, we will conclusively measure the abundances of galactic-scale quasar and AGN pairs, offset AGNs, and sub-arcsec lensed quasars across most of the cosmic history, with unprecedented statistics and coverage of the parameter space of SMBHs and host galaxies. In turn, these data will deliver an observational consensus on the formation and evolution of SMBH pairs in galactic environments. \\

\noindent\fire{Technical Aspects}\quad This science case rests on the High Latitude Wide Area Survey (HLWA) with Roman's Wide Field Instrument. The current baseline design of the HLWA survey will cover $>1700\,{\rm deg^2}$ in 4 NIR bands (0.93--2.00\,\micron), reaching a $5\sigma$ imaging depth of $\sim 26-27$ in $YJH$ and F184. This sensitivity and area coverage will probe AGN luminosities down to $L_{\rm bol}\gtrsim 10^{42}\,{\rm erg\,s^{-1}}$ at $z>1$ (i.e., a factor of $10^4$ fainter than luminous SDSS quasars) and uncover more than $\sim 10$ million AGNs over $1<z<7$ (the exact number is unknown without actual measurements of the AGN LF at the Roman magnitude limit). While theoretical uncertainties are still large, recent simulations generally predict a galactic-scale dual AGN fraction of $\sim 1\%$ among all AGNs with $L_{\rm bol}>10^{43}\,{\rm erg\,s^{-1}}$ at $z>1$, and ten times more abundant single offset AGNs in galaxy mergers \citep[][]{DeRosa_etal_2019}. Thus we expect $>10^5$ ($>10^6$) dual (offset) AGNs on galactic scales from the HLWA survey. These sample sizes linearly scale with the survey area, assuming fixed depth. These unprecedented statistics are sufficient to bin the dual/offset AGN sample in pair separation, redshift, AGN luminosity and host galaxy types, allowing us to disentangle different physical mechanisms that drive the formation and evolution of SMBH pairs. 

While we do not require significant changes in the HLWA design to enable dual SMBH science, we emphasize on the following technical requirements for Roman data processing and policies: (1) To enable a complete coverage of pair statistics down to the $\sim 0.13$" resolution of Roman, it is important to deblend these close pairs (often within extended host galaxies or mergers) efficiently and robustly. Although it is possible to quantify and correct for the incompleteness of pair resolving, it is crucial for Roman to deliver the best possible deblended photometry. A secondary requirement is that deblending should retain a high success rate for pair contrasts of as low as 0.1 or 0.3 to reduce the bias of preferentially selecting equal-flux pairs. (2) While the Roman grism spectroscopy will undoubtedly be useful for our science, the nature of pair proximity and required spectral sensitivity may limit the utility of these grism spectra beyond a redshift determination. Instead, our science case will heavily rely on robust photometric identification of unobscured AGNs. As such, it is critical to combine NIR photometry from Roman with optical photometry from other space-based wide-area imaging surveys.

\clearpage

\bibliography{refs}

\end{document}